# Contact line friction of bubbles

Xicheng Bao[1,2], Aaron D. Ratschow[1], Xiaoteng Zhou[3], Chirag Hinduja[1], Xiaomei Li[4], Qinshan Liu[2], Dandan Gao[5], Xiahui Gui[2], Rüdiger Berger[1], Yaowen Xing[2*], Hans-Jürgen Butt[1*], Michael Kappl[1*]

[1] *Max Planck Institute for Polymer Research, Ackermannweg 10, 55128 Mainz, Germany*

[2] *State Key Laboratory of Coking Coal Resources Green Exploitation, China University of Mining and Technology, Daxue Road 1, 221116 Xuzhou, China*

[3] *Department of Mechanical Engineering, Massachusetts Institute of Technology, Cambridge, MA 02139, USA*

[4] *Department of Chemistry and Applied Biosciences, ETH Zurich, 8093 Zürich, Switzerland*

[5] *Department of Chemistry, Johannes Gutenberg University, Duesbergweg 10-14, 55122 Mainz, Germany*

E-mail: kappl@mpip-mainz.mpg.de

**Abstract:**

Contact line friction (CLF) of bubbles is ubiquitous, from bubbles on a beer glass to $H_2$ bubbles sliding over electrodes in electrolysis. However, a fundamental understanding of CLF of bubbles is still missing, mainly due to the challenge of precisely controlling bubble sliding. For example, it is not clear how bubbles start sliding and how CLF of bubbles depends on velocity. We therefore developed a bubble friction force instrument to directly measure bubble CLF. This force develops from a static regime, through a transition, to a kinetic regime. This entire process is quantitatively described by a modified Kawasaki-Furmidge equation. Bubble CLF was measured for velocities from 0.2 μm/s to 2 mm/s, revealing a transition from a constant CLF regime below about 60 μm/s to a velocity-dependent CLF regime on surfaces with various wettability. The velocity dependence stems from interfacial adaptation governed by the liquid ionic environment with a relaxation time of around 10 μs. Moreover, CLF of bubbles can be measured on hydrophilic surfaces and under a challenging $H_2$ atmosphere, overcoming the limitations of current droplet-based methods. Our results provide a quantitative basis for understanding CLF of bubbles with relevance to many applications, including bubble manipulation and electrochemistry.

The walls of a freshly filled beer glass are commonly covered with countless bubbles—a familiar sight that conceals an intriguing physical question: Where does the friction resisting buoyancy originate? Friction force at the solid-vapor interface within the bubble contact area is negligible; thus, the friction arises from the contact line, and is referred to as contact line friction (CLF)[1,2]. Every wetting and dewetting process fundamentally hinges on the contact line movement[3], which is strongly governed by CLF opposing the driving force at contact line[4]. Beyond everyday phenomena, CLF is ubiquitous in broad applications and especially in bubble-related fields[5–8]. For example, in electrochemical processes, the retention of bubbles can block electrode surfaces, resulting in a 10-25% decrease in electrode performance[9]. Bubble CLF serves as a fundamental mechanical basis for directed bubble transport and microfluidic applications[10]. Moreover, flotation, which is the primary beneficiation method for rare-earth minerals, depends critically on the control of bubble CLF to ensure successful separation[11]. An in-depth understanding of bubble CLF is essential for effectively controlling such systems.

Current research on CLF predominantly relies on Wilhelmy plate method[12] and droplet-based methods (e.g., tilted plate method and drop adhesion force instrument)[13,14]. The Wilhelmy plate method quantifies the velocity dependence of capillary force $\gamma_{LV} \cos\theta_{adv}$ or $\gamma_{LV} \cos\theta_{rec}$ [15]. Here, $\gamma_{LV}$ is the liquid's surface tension, $\theta_{adv}$ and $\theta_{rec}$ are the advancing and receding contact angle, respectively. CLF would be $\gamma_{LV}(\cos\theta_e - \cos\theta_{adv})$ or $\gamma_{LV}(\cos\theta_{rec} - \cos\theta_e)$. But, as there is no practical method of measuring "equilibrium" contact angle $\theta_e$, the contribution of CLF cannot be separated from the capillary force. In 1950, the tilted plate method was introduced, where droplets are driven by gravity, to link contact angle hysteresis (CAH) with droplet sliding friction[16]. This approach led to the well-known Kawasaki-Furmidge equation[17,18], which can predict the force required to slide a drop over a surface based on its interfacial morphology. In 2012, drop adhesion force instrument (DAFI)[19] was developed, which employs a motor-driven surface to slide against the droplet at a constant velocity and measure the droplet friction directly via a force sensor[20–22]. Remarkably, the friction curves for droplets on surfaces were found to resemble solid-solid friction curves (Fig. 1a, b)[4]. Bubbles and droplets exhibit fundamental differences, including their phase configuration, contact angle positions, and vapor pressure. Extending the knowledge of CLF from droplets to bubble systems faces fundamental challenges[23]. However, the CLF of bubbles has remained largely unquantified. Measurements to date are confined to static or quasistatic bubbles on hydrophobic surfaces[24,25]. The quantitative measurement of CLF of sliding bubbles has remained challenging due to the difficulty of precisely manipulating a bubble for a friction force test. In existing studies of buoyancy-driven sliding bubbles on inclined plates, the direct quantification of bubble CLF is prevented by a complex coupling of effects, especially the non-constant sliding velocity[26–28]. Therefore, it is still not clear how bubbles start sliding and how CLF of bubbles depends on velocity. This gap calls for an approach that directly quantifies bubble CLF to reveal its fundamental nature.

For general research on air-water interface CLF, the droplet-based methods share some limitations. Due to the droplet breakup and the influence of the precursor film[29], droplet-based methods may fail when applied to hydrophilic surfaces. Additionally, evaporation of volatile liquids prevents low-speed CLF measurements, making it difficult to study the origin of its velocity dependence[4]. Furthermore, it is challenging to switch atmospheres like $H_2$ for studying CLF. Given these challenges, could a bubble friction method overcome the limitations of current methods? What new insights into contact line physics can the study of bubble CLF provide?

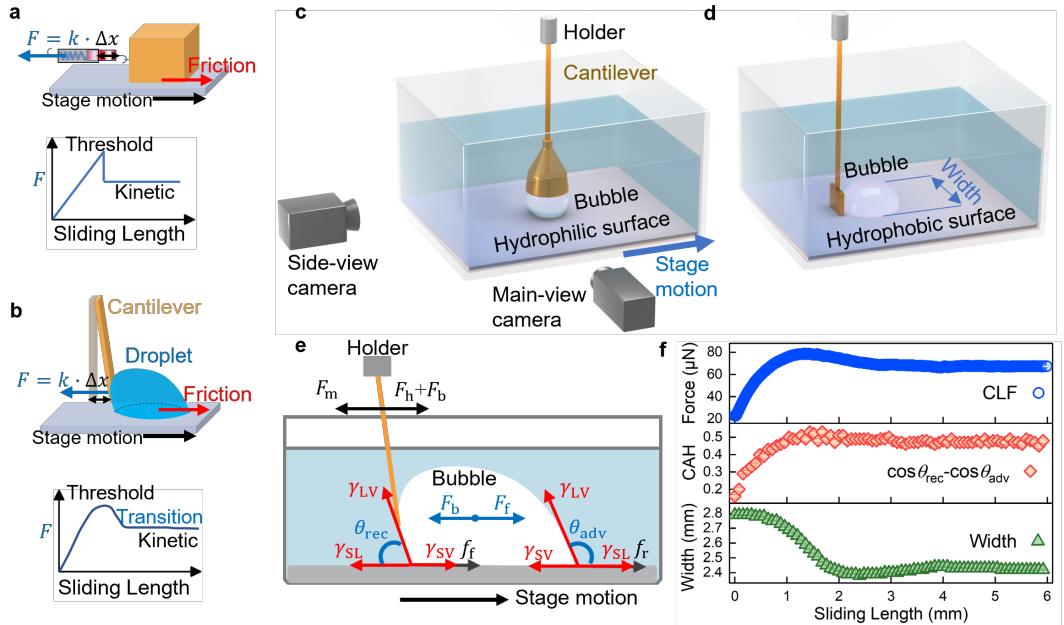

**Figure 1 | Schematics of CLF measurements. a, b,** Setup and force curve of solid–solid friction (**a**) and droplet–solid friction (**b**). **c, d,** BuFFI for hydrophilic (**c**) and hydrophobic surfaces (**d**). **e,** Forces acting on the cantilever and bubble: cantilever's deflection force $F_m$, hydrodynamic drag force acting on both the bubble and the cantilever $F_h$, bubble–cantilever interaction force (excluding hydrodynamic drag, already accounted for in $F_h$) $F_b$, bubble friction force exerted by the substrate surface $F_f$, CLF of the contact line front-end $f_f$ and rear-end $f_r$. **f**, Typical measurement results at a velocity of 0.2 mm/s on a fluorinated silicon wafer for CLF, contact angle hysteresis, and width. Error bars indicate standard deviations determined from three independent measurements. Here, the errors are smaller than the symbol size due to the scale of the y-axis.

To address these questions, we have developed the bubble friction force instrument (BuFFI). A submerged gas bubble is attached to a fixed cantilever while a liquid tank with the substrate moves at a constant velocity via a positioning stage (Fig. 1b-d). The design details of cantilever springs are provided in Supplementary Section 1. Through this relative movement, the bubble slides along the substrate. Using Hooke's law, the cantilever's deflection force, $F_m$, is calculated from the measured deflection and the cantilever spring constant (Supplementary Section 2). All forces are treated as scalars, and directions are indicated by arrows in the schematic (Fig. 1e). For cantilever, in steady state, $F_m$ can be written as a sum of two contribution: $F_m = F_h + F_b$. The hydrodynamic drag force $F_h$ is acting on both the bubble and the cantilever. It can be determined by moving the platform with the bubble close to the surface but without direct bubble/surface contact (Supplementary Section 3). $F_b$ is the interaction force between bubble and cantilever. It excludes hydrodynamic drag force, which is already accounted for in $F_h$ above. It can be easily calculated by $F_b = F_m - F_h$. For bubble, $F_b$ is balanced by the bubble friction force $F_f$ exerted by the substrate surface. Thus, the bubble friction force can be calculated by $F_f = F_b$. During bubble sliding, the induced fluid flow within the liquid tank generates viscous dissipation[30]. Viscous dissipation in liquid near the moving contact line of the sliding bubble contributes to $F_b$ and is a part of CLF[31]. Viscous dissipation at the substrate solid-liquid interface far from the contact line does not contribute to cantilever deflection and viscous dissipation at the solid-gas interface is negligible due to the low gas viscosity. Thus, the bubble friction force $F_f$ is equivalent to the CLF (Fig. 1e). BuFFI can therefore directly quantify CLF. Within the velocity range in this study, the hydrodynamic drag force $F_h$ is within the noise floor of the force measurement and can be neglected (Supplementary Fig. 4). The measured deflection force $F_m$ directly quantifies CLF (Fig. 1f).

Unlike droplet-based methods, the BuFFI allows for CLF measurements on both hydrophobic and hydrophilic substrates. On hydrophilic substrates, we generate a bubble through a custom-made, hollow cantilever (Fig. 1c). The shape of the cantilever keeps the bubble in place, as it would easily detach from hydrophilic substrates. For hydrophobic substrates, we generate a bubble on the surface using a micro syringe and then laterally attach it to the end the cantilever (Fig. 1d).

The CLF of bubbles shows both a distinct threshold force and a transition regime (Fig. 1b, f). The features resemble the friction force of sliding droplets[4]. The CAH trend closely follows the CLF (Fig. 1f), indicating a strong association between CAH and

5 / 24

CLF. In this paper, we define CAH as $\cos\theta_{\text{rec}} - \cos\theta_{\text{adv}}$. In contrast, the width of the contact line lags behind the changes of CLF and CAH, because the contact line begins to move gradually from front to rear, rather than all at once after threshold. The bubble's front contact line moves first, and this motion then propagates to the bubble's rear contact line (Supplementary Section 4). It is commonly assumed that the region before the threshold corresponds to a static friction regime, following the classic solid–solid friction law, and many studies on droplet sliding friction use the same division. For consistency and easy comparison, we also refer to the pre-threshold region as the "static" friction regime of CLF.

We directly measured CLF on a cleaned Si wafer (see Methods) as a hydrophilic substrate with static advancing and receding contact angles of 54° and 26°, respectively (Fig. 2a). As the moving distance of the BuFFI platform increases, the force first increases linearly due to increasing cantilever bending while the contact line of the deforming bubble is still at rest. After reaching a threshold force, it decreases through a transition regime into a stable kinetic regime. This force behavior differs from that of droplet friction on such a hydrophilic surface. Sliding droplets will strongly be stretched out on the receding side due to the low receding contact angle on hydrophilic surfaces, resulting in droplet breakup. The breakup events lead to additional capillary forces, resulting in jumps in the force curve, as evident in the results of a drop sliding experiment carried out by us on the same Si wafer (Fig. 2b).

We also measured CLF on two other surfaces with higher contact angles: on indium tin oxide coated glass (ITO-glass) with static advancing/receding contact angles of 65°/29° and on fluorinated Si wafers with static advancing/receding contact angles of 110°/95° (Fig. 2c, d). Therefore, bubble CLF is measurable not only on hydrophobic but also on hydrophilic surfaces where droplet-based methods tend to fail.

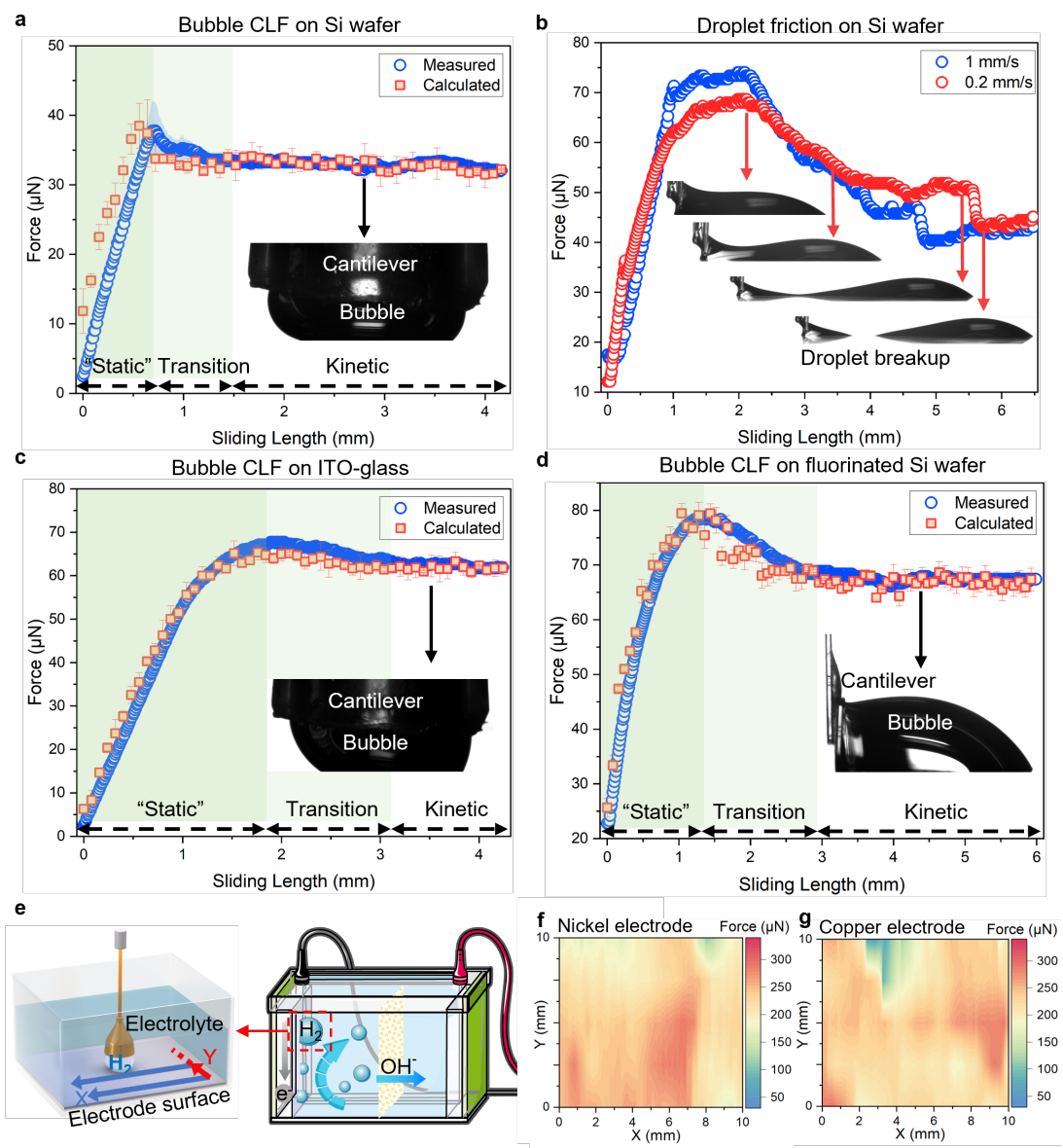

**Figure 2 | CLF measurement in different situations. a,** CLF traces for a 10 μL bubble sliding at a velocity of 0.2 mm/s over a hydrophilic Si wafer. Blue circles denote measured friction forces. Red squares indicate forces calculated from measured contact angles using equation (1) with $k$=0.72. Bubble images (insets) were captured during the kinetic regime as indicated by arrows. The initial force exceeds 0 due to the attractive interaction between the cantilever and the bubble during attachment. Error bars indicate standard deviations determined from three independent measurements. **b,** Droplet sliding on Si wafer leads to splitting and jumps in the CLF trace. **c, d,** CLF traces and fitting results for ITO-glass with $k$=0.99 (**c**) and fluorinated Si wafer with $k$=0.81 (**d**). **e**, Schematics of creating 2D CLF maps by line-wise scanning on an electrode surface

with a 10 μL H$_2$ bubble in the electrolyte. **f, g,** 2D CLF map of Nickel electrode (**f**) and Copper electrode (**g**). Each scan has a sliding length of 12 mm in the x-axis, so that the kinetic regime of every scan was at least 10 mm long. 5 scans are performed with the scanning stage being advanced by 2 mm in the y-axis after each scan in x-direction. The raw force data are shown in Supplementary Fig. 8. Using a linear interpolation algorithm, all the raw force curves were combined into a continuous 2D CLF contour map.

The CLF acting on a bubble that is static or about to slide is given by integrating the lateral capillary forces per unit length around the contact line:

$$F_c = 2\int_0^\pi \gamma_{LV} \cos\theta^b \cdot \cos\phi \cdot y d\phi = kw\gamma_{LV}(\cos\theta^b_{rec} - \cos\theta^b_{adv}) \qquad (1)$$

Here, $\phi$ is the azimuthal angle, $y(\phi)$ is the distance of the contact line to the axis of motion (Supplementary Section 7), $k$ is shape factor determined by the distribution of the contact angle along the contact line, $w$ is the width of the contact area (Fig. 1d), $\theta^b(\phi)$ is the local bubble contact angle measured through the vapor phase. The form of equation (1) is commonly referred to as the Kawasaki-Furmidge equation. We use here $y(\phi)$ instead of the traditional contact line radius $\xi$[32], since it is the physically appropriate quantity (Supplementary Section 7). We therefore term equation (1) the modified Kawasaki-Furmidge equation. To remain consistent with the conventional definition, we use contact angles measured through the liquid phase. Thus, the term in brackets becomes $\cos(\pi - \theta_{adv}) - \cos(\pi - \theta_{rec})$, which is equal to $\cos\theta_{rec} - \cos\theta_{adv}$.

In the kinetic regime, the bubble deforms as it slides, which in turn might alter $k$. The fitting of bubble CLF results with equation (1) yielded good agreement between calculated and measured CLF (Fig. 2a, c, d). Notably, the calculated CLF reproduced not only the static regimes but also the transition and kinetic regimes, which indicates that $k$ does not change significantly. Under defined experimental conditions, we obtained $k = 0.72$ for the Si wafer, $k = 0.99$ for ITO-glass, and $k = 0.81$ for the fluorinated Si wafer. Thus, the CLF of bubbles, from the static to the kinetic regime, can be well described by the modified Kawasaki-Furmidge equation.

Bubble CLF can be easily quantified under defined gas atmospheres. In contrast, droplet-based methods for studying CLF would typically require a gas-tight sealed chamber with a controlled atmosphere. For gases such as $H_2$, it becomes challenging due to safety constraints[33]. As an example, we applied BuFFI to electrode surfaces used for water splitting to produce $H_2$. When conducting electrochemical linear sweep voltammetry (LSV, Supplementary Fig. 6) on two commonly used electrodes, nickel and copper, the primary bubble behavior at the electrode surfaces includes bubble growth, sliding, and detachment, all of which involve contact line motion. Bubble CLF can therefore provide a direct and quantitative link to these steps, especially during the sliding and detachment[6].

As a principle proof of using BuFFI to study CLF for hydrogen on electrode surfaces, we performed line-wise scanning[34] using hydrogen bubbles on electrode surfaces (Fig. 2e). CLF scanning was also performed on an ITO-glass surface (Supplementary Fig. 7). The CLF scanning results reveal significant fluctuations in CLF on real electrode surfaces, ranging from approximately 50-300 μN (Fig. 2f, g). These fluctuations are attributable to surface roughness (Supplementary Fig. 10c). Under defined gas and solution conditions, such CLF scans yield CLF contour maps that can help in optimization of electrode surface designs[6]. Scanning directly provides bubble CLF data without the need of empirical parameters, thereby minimizing potential errors[35].

CLF of bubbles can be measured at very low velocities, as bubbles maintain stable volume and vapor pressure. By attaching a bubble to the cantilever and recording its height over time, we found that the bubble profile remained nearly unchanged for 30 minutes, and its height decreased by only 0.07 mm after 12 h due to the low dissolution rate of the gas (Fig. 3a). The gas inside the bubble is automatically 100% saturated with water vapor. In contrast, sessile water droplets evaporate under atmospheric conditions, as the vapor pressure is below 100%. For example, a 5 µL drop at 20 °C temperature and 49% humidity evaporated in 37 min (Fig. 3b). Consequently, droplet friction at a low velocity (6 µm/s) is strongly influenced by drop evaporation, leading to a continuous decline of the droplet friction due to the shrinking droplet size (Fig. 3c). By contrast, at a very low velocity (0.2 µm/s), bubble CLF shows a stable kinetic regime (Fig. 3c). The theoretical minimum operating speed can be as low as nm/s, depending on the motor's speed limit.

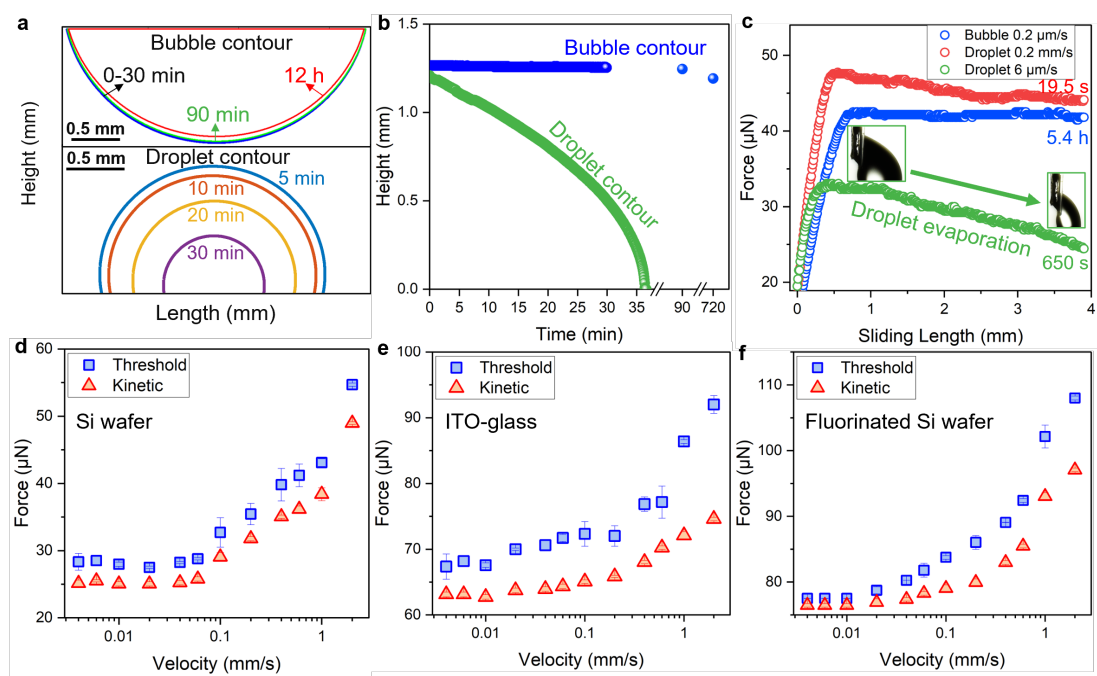

**Figure 3 | Contour evolution of bubble and droplet and Velocity dependence of CLF. a**, The contour variation with time of a 10 µL air bubble attached to the cantilever used for hydrophilic surfaces underwater and a 5 µL droplet on fluorinated Si wafer in open air. **b**, The contour height of bubble and droplet versus time. **c**, Evaporation effect of droplet friction and bubble CLF on fluorinated Si wafer with a 5 µL droplet and bubble. **d, e, f,** CLF versus velocities on Si wafer (**d**), ITO-glass (**e**) and fluorinated Si wafer (**f**). Velocities range from 4 µm/s to 2 mm/s. The kinetic CLF is determined from the constant-force segment within the kinetic regime, and the forces of different velocities are selected at the same bubble sliding location. Error bars indicate standard deviations determined from three independent measurements.

We measured CLF on a Si wafer, ITO-glass, and a fluorinated Si wafer for velocities ranging from 4 μm/s to 2 mm/s. In all cases, for low velocity CLF is speed independent up to a distinct turning point. Beyond the turning point, CLF increases with velocity. Both the threshold force and kinetic force follow similar trends (Fig. 3d, e, f). The turning point at which velocity dependence in CLF sets in is ≈60 μm/s. To interpret this velocity dependence, we separately evaluated the applicability of hydrodynamic viscous dissipation[36], molecular kinetic theory (MKT[37,38]) and adaptation theory[39] within this velocity range.

To exclude the influence of viscous dissipation, we measured CLF in water (viscosity η = 1 cP) and in 20 vol% glycerol aqueous solution (viscosity η = 2 cP) on Si wafers. For better comparison, we plot their CAH versus velocity (Supplementary Fig. 11). According to equation (1), CAH equals the scaled CLF $\cos\theta_{\text{rec}} - \cos\theta_{\text{adv}} = F_f/(kL\gamma)$. We find negligible differences. We conclude that viscous dissipation due to shear in water is negligible within our velocity range below 2 mm/s. MKT, in which the motion of the contact line is modeled as a molecular adsorption-desorption process involving a series of small jumps, did not fit our force-versus-velocity plots (Supplementary Section 10).

To interpret the velocity dependence, we developed an interfacial adaptation model for sliding bubbles (Supplementary Section 11). In all theories describing wetting and wetting dynamics, it is assumed that the solid surface energy $\gamma_{\text{SV}}$ immediately changes to the solid-liquid interfacial energy $\gamma_{\text{SL}}$ when the liquid got in contact with the solid (Fig. 4a, bottom schematic). However, when a real solid surface is immersed in water, liquid molecules may diffuse into it (especially for polymers) (Fig. 4b), surface groups may reconstruct (Fig. 4c), an electric double layer (EDL) is formed (Fig. 4d), liquid molecules reorient (Fig. 4e), etc. All these adaptation processes take time[39]. The interfacial tensions change from an initial value $\gamma_{\text{SL}}^0$ to a new equilibrium value $\gamma_{\text{SL}}^\infty$, which is reached when enough time has passed for the adaptation processes to fully occur (Fig. 4a, middle schematic). Here, we assume that $\gamma_{SL}^0$ is higher than $\gamma_{SL}^\infty$ because the relaxation is spontaneous. The same applies to the rear contact line.

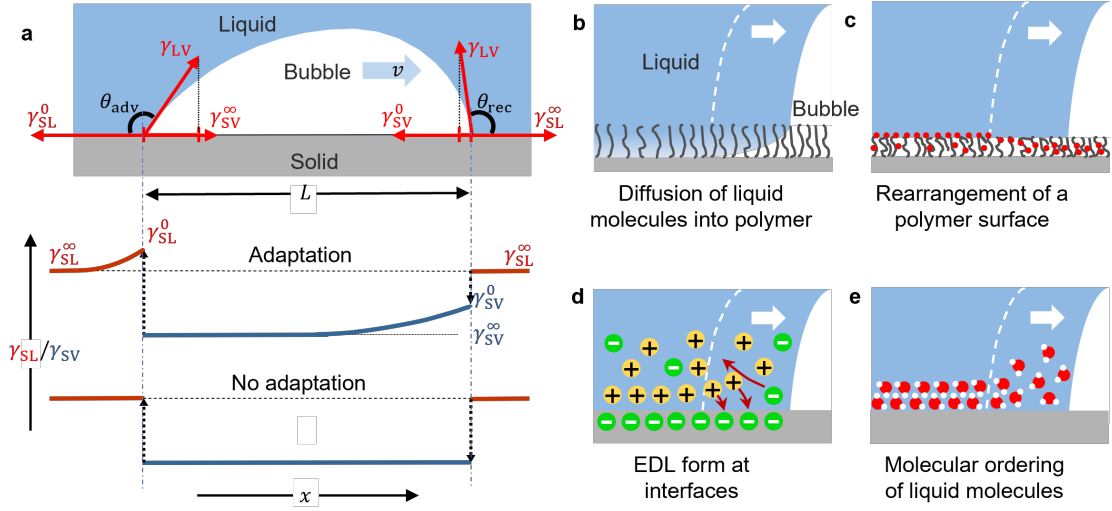

**Figure 4 | Schematic of interfacial adaptation of a moving bubble and examples of adaptation process. a**, Position-dependent interfacial tension development for a bubble in steady-state sliding on a surface, with and without adaptation considered. b, Diffusion of liquid molecules into polymer. Black chains represent polymer molecules. The white arrow represents the moving direction of the bubble. The interior of the white dashed line has been fully adapted. c, Rearrangement of a polymer surface. Lyophilic side groups (red circles) generally point toward the liquid. d, EDL dynamics at interfaces. e, Molecular ordering of liquid molecules.

Based on the solid-liquid interfacial energy relaxation during bubble sliding, an expression for the advancing contact angle $\theta_{adv}$ that depends on velocity is derived:

$$\cos\theta_{adv} = \cos\theta_{adv}^\infty - \frac{\Delta\gamma_{SL}}{\gamma_{LV}^\infty} e^{-v_{SL}/v} \quad (2)$$

Here, $\theta_{adv}^\infty$ is the advancing contact angle in the infinitely slow and steady-state advancement of the contact line, $\Delta\gamma_{SL} = \gamma_{SL}^0 - \gamma_{SL}^\infty$, $v_{SL}$ is the adaptation velocity, $v$ is the actual contact line velocity. The adaptation velocity is defined by $v_{SL} = l/\tau_{SL}$, where $\tau_{SL}$ denotes the adaptation relaxation time at the solid-liquid interface and $l$ is the contact line core region width, which influences the contact angle[39,40].

Similarly, when the liquid recedes and the solid is re-exposed to vapor, the expression for the velocity dependence of $\theta_{\text{rec}}$ is:

$$\cos\theta_{\text{rec}} = \cos\theta_{\text{rec}}^{\infty} + \frac{\Delta\gamma_{\text{SV}}}{\gamma_{\text{LV}}^{\infty}}e^{-v_{\text{SV}}/v} \tag{3}$$

Here, the parameters are defined similarly, but for the solid-vapor interface.

We applied the model to fit the advancing and receding contact angles on three different surfaces. Adaptation theory yielded accurate fitting results on all three surfaces, with the fitted curves closely matching the turning points of the force data (Fig. 5).

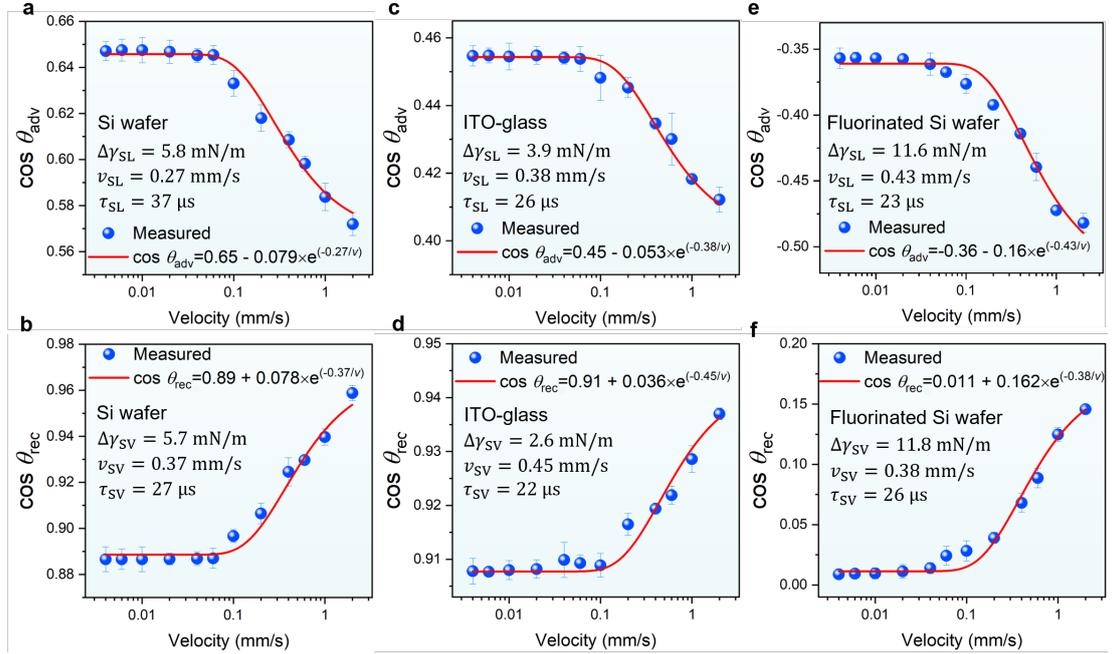

**Figure 5 | The velocity dependence of advancing and receding contact angle fitted by adaptation theory.** Adaptation theory fitted $\cos\theta_{\text{adv}}$ and $\cos\theta_{\text{rec}}$ in the kinetic regime on Si wafer (**a**)(**b**), ITO-glass (**c**)(**d**) and fluorinated Si wafer (**e**)(**f**). The contact angle measurement position is identical to the force selecting position in Fig. 3d,e,f, and the error bars indicate standard deviations.

Taking the Si wafer as an example, we illustrate the origin of velocity dependence as explained by adaptation theory. The adaptation velocity of the adaptation process on the Si wafer surface was $v_{\text{SL}}$=0.27 mm/s and $v_{\text{SV}}$=0.37 mm/s. At very low velocities ($v \ll v_{\text{SL}}$), the adaptation process within the peripheral thickness is fully completed, allowing the $\cos\theta_{adv}$ and $\cos\theta_{rec}$ to reach their equilibrium values, $\cos\theta_{adv}^{\infty} = 0.65$ and $\cos\theta_{rec}^{\infty} = 0.89$. The CAH remains constant at $\cos\theta_{rec}^{\infty} - \cos\theta_{adv}^{\infty} = 0.24$. According to equation (1), bubble friction is constant. However, as the velocity increases beyond the turning point in the force–velocity relationship, $\theta_{adv}$ and $\theta_{rec}$ no longer have sufficient time to reach their equilibrium values. Consequently, with increasing speed, $\cos\theta_{adv}$ starts to decline and $\cos\theta_{rec}$ begins to rise, resulting in increased CAH and causing the CLF force to steadily increase. These changes explain the origin of the observed velocity dependence.

Based on the fitting results, we calculated $\Delta\gamma_{SL}$, $v_{SL}$, $\Delta\gamma_{SV}$, $v_{SV}$ for each surface (Fig. 5). The relaxation time is determined by the adaptation velocity and peripheral thickness, which depends on the specific nature of the interface. A typical estimate for the peripheral thickness is 10 nm, based on the range of surface forces and the solid region experiencing stress[39]. The relaxation times can be calculated based on the observed adaptation velocity and the assumed peripheral thickness of 10 nm. For the advancing side, the adaptation relaxation time is $\tau_{SL}$=37 μs on Si wafer, 26 μs on ITO–glass, and 23 μs on fluorinated Si wafer, all of similar magnitude. A similar result was obtained for the receding side. Therefore, the experimental results indicate that moving contact lines on these surfaces with different wettability involves an adaptation process with a typical relaxation time scale of ~10 μs.

This observation indicates that the adaptation process is governed by liquid molecules behavior rather than the specific surface properties of the solid. Since the reorientation of water molecule is known to occur on the picosecond time scale[41], it is unlikely. It is more reasonable to attribute the observed adaptation to the EDL dynamics. At the advancing contact line, an EDL forms at the solid-liquid interface with a possible influence of the existing EDL at the liquid-vapor interface. At the receding contact line, it is disturbed by the upward flow near the contact line and the reformation of the EDL at the liquid-vapor interface[42]. In most cases, EDL dynamics is limited by the transport of ions towards and away from the interface. The corresponding relaxation time can be estimated by the Debye time[43], $\tau_D = \lambda_D^2/D$, where $\lambda_D$ is the Debye screening length and $D$ is the ion diffusivity across $\lambda_D$. We measured the pH of the ultrapure water that we used. It was around 6 due to absorbed carbon dioxide[44,45]. With a mean diffusivity of hydroxyl and hydronium ions[46] of $D \approx 7 \times 10^{-9}$ m²/s and a Debye screening length[47] of $\lambda_D \approx 300$ nm we estimate a relaxation time of the EDL of $\tau_D \approx 13$ μs. This value is of the same order as the relaxation time of the adaptation process, confirming that the interfacial adaptation might be governed by EDL dynamics.

To validate the hypothesis, we changed the ion concentration using NaCl solutions at 0.1 mM, 1 mM, and 10 mM. The adaptation relaxation time is $\tau_{SL}$=0.6 μs on Si wafer and 0.8 μs on fluorinated Si wafer at 0.1 mM. Similar results were obtained for the receding side. In theory, the predicted relaxation time of EDL dynamics at 0.1 mM is 0.6 μs. The results for 0.1 mM NaCl solutions support the hypothesis, whereas those for 1 mM and 10 mM deviate from it (Supplementary Section 12). It suggests that at low ion concentrations (≤0.1 mM), including ultrapure water, the adaptation molecular process is governed by EDL dynamics. At higher ion concentrations a more complex adaptation process appears to increase the relaxation time, the underlying mechanism remains unclear.

In this work, we present a method that directly quantifies bubble CLF across diverse conditions and provide quantitative models for the behavior and velocity dependence of bubble CLF. BuFFI opens new avenues in contact line physics by overcoming the limitations of current methods. It unlocks quantitative studies of CLF on hydrophilic surfaces, under specific atmospheres like H₂, and with ultra-low speed. More generally, our work will foster research both on fundamental and applied aspects of bubble and contact line physics, including bubble manipulation, bubble sliding electrification, and microfluidic routing.

## Methods

### Sample preparation

Three distinct surface types were utilized in this study: Pristine Si wafer (Type/Dopant: P/B; Thickness: 675 ± 25 µm; Siegert Wafer), fully oxidized indium tin oxide (ITO) glass substrates (Dimensions: 25 × 75 mm$^2$; Thickness: 1.1 mm; Supplier: Ossila), and fluorinated Si wafer. The fluorination process was performed as follows: pristine Si wafer was sequentially ultrasonicated in toluene (99.99%, Fisher Chemical), acetone (99.98%, Fisher Chemical), and absolute ethanol (99.96% v/v; assay on anhydrous substance, 100.0%; VWR Chemicals) for 10 minutes each. After cleaning, the wafer underwent $O_2$-plasma treatment at 300 W for 10 minutes using a Femto low-pressure plasma system (Diener Electronic). Subsequently, the Si wafer was placed inside a vacuum desiccator containing 20 µL of 1H,1H,2H,2H-perfluorodecyl-trichlorosilane (PFDTS, 96%, Alfa Aesar). The desiccator was evacuated to pressures below 100 mbar, sealed, and maintained for a reaction period of 120 minutes. Prior to measurements, the PFDTS-coated surfaces were ultrasonicated in isopropanol (99.8%, Honeywell) for 5 minutes to remove any unbound silane molecules. Pristine Si wafer and ITO-glass were sequentially ultrasonicated in toluene, acetone, and absolute ethanol for 10 minutes each before use. Representative scanning electron microscope (SEM) images are shown in Supplementary Fig. 10.

### Measurement of static advancing and receding contact angles

The static advancing and receding contact angles were quantified using the Drop Shape Analyzer (DSA100, KRÜSS) with the method of in-/deflated sessile water droplets. The procedure was initiated by depositing a 4 µL liquid drop onto the tested surfaces. Following this, a Hamilton syringe equipped with a hydrophobic needle pumped 30 µL of liquid into and then out of the drop at a flow rate of 1 µL/s. This process was carried out three times without pausing at one area, and measurements were performed at three different areas. Throughout the procedure, the inflation and deflation of the drop were recorded from the side. The static advancing and receding contact angles were determined through a tangent fitting to the contour of the drop.

**Spring constant of cantilevers**

The spring constants of the cantilevers were calibrated by correlating the applied force with lateral deflection[4]. The main body of the cantilever is a rectangular glass capillary (50 × 0.5 × 0.05 mm3, VitroCom's Vitrotubes, CM Scientific). Initially, one end of the capillary was secured in a copper holder. This assembly was then attached to a micromanipulator (MMO-203, Narishige), enabling precise movements with 1 μm accuracy. A pin, equipped with supporting pedestals and a pointed tip, was placed on a microgram balance (Supplementary Fig. 2). Using the micromanipulator, the free ends of the cantilevers were positioned directly above the pin tip. The cantilevers were gradually lowered until they contacted the pin, then restored to a zero-deflection position to offset the deflection caused by their own weight. The contact position was recorded for subsequent force measurements based on deflection. The microgram balance measured the force exerted between the pin and the deflected cantilevers. Supplementary Fig. 3 illustrates a typical calibration curve for the cantilevers' spring constant. According to Hooke's law, the spring constant was determined from the slope of the linear fit.

**Force measurements**

All experiments were conducted at a temperature of $20 \pm 1$ °C and a humidity of 45–50% using Milli-Q ultrapure water with a typical resistivity of 18.2 MΩ·cm. A stepper motor (4H4018L0502, TECO) enables precise movement of the liquid tank in the bubble sliding direction, defined as the X direction. During CLF scanning, a separate motor (PS3210, KRÜSS) controls the movement of the liquid tank along the Y direction. Bubbles were generated using a 25 μL micro-syringe. For the hydrophilic surface setup, a 10 μL bubble was generated at the tail end of the cantilever, while for the hydrophobic surface setup, a 5 μL bubble was placed directly on the surface. Using different bubble sizes can make the initial contact line width of the two setups similar.

Prior to initiating measurements, the XY coordinates of the stage were recorded as the starting point. The position of the cantilever was monitored using a main-view camera (UI-3060CP-M-HQ R2, IDS), with the force measurement position corresponding to the cantilever tip during spring constant calibration. Deflection of the cantilever at the force measurement position was calculated using MATLAB R2022b software, and Hooke's law was applied to determine the force at various sliding lengths. The relative error of this method is 1.85% for the cantilever used for hydrophilic surfaces and 1.84% for the cantilever used for hydrophobic surfaces, respectively. (Supplementary Fig. 15)

For parallel experiments and tests conducted at different speeds on the same surface, the instrument can return the bubble to the pre-recorded starting point, thereby eliminating force discrepancies caused by variations in surface properties across different regions and ensuring controlled variables. Viscosity measurements were performed using 20 vol% glycerol (≥99.5%; Sigma-Aldrich) aqueous solutions, with the starting point consistently set to the pre-recorded position.

CLF measurements of hydrogen bubbles on copper (Thickness: $0.2 \pm 0.02$ mm; Metall Ehrnsberger) and nickel (Thickness: $0.2 \pm 0.02$ mm; Metall Ehrnsberger) electrode surfaces were conducted in Ar purged 0.1 M phosphate-buffered saline electrolyte (pH 7.4). Hydrogen gas was pre-stored in clean balloons designated for chemical experiments and introduced as hydrogen bubbles at the hydrophilic cantilever's end via a conduit.

**Bubble imaging**

During the bubble sliding process, the main-view camera (UI-3060CP-M-HQ R2, IDS) simultaneously recorded the advancing and receding contact angles. A side-view camera (FASTCAM Mini UX100 type 800K-M-8G, Photron) captured the bubble width throughout the sliding motion. Both contact angles and bubble widths were measured using ImageJ software. Synchronization between the side-view and main-view cameras was achieved through triggering.

**SEM Imaging**

Images were acquired using a Hitachi SU8000 SEM operated at an acceleration voltage of 3 kV. Secondary electron signals were collected. All samples were sputtered with Pt using BalTec MED 020 modular high vacuum coating system (argon at $2\times10^5$ bar, 5 nm).

**Acknowledgements**

We thank Y. Xiang, Z. Ni, H. Luo, H. Yang and W. Steffen for their valuable suggestions, and Y. Dong, S. Azadeh and G. Schäfer for technical support. This work is supported by the Sino-German Mobility Programme 'New principles of flotation separation based on surface/interfacial micro-nano mechanics' (Grant No. M-0230, Y.X. and X.G.), the German Research Foundation (DFG) within the framework Collaborative Research Centre 1194 'Interaction of Transport and Wetting Processes', Project-ID 265191195, subproject C07 and T02 (C.H., R.B. and H.-J.B.) and the Priority Program 2171 'Dynamic wetting of flexible, adaptive, and switchable surfaces' (Grant No. BE 3286/6-1, R.B.). X.Z. thanks for the support from the DFG with grant No. 550194666. D.G. gratefully acknowledges the DFG for a Walter Benjamin Fellowship (project no. 510966757), the financial support by the Carl Zeiss Foundation (Halocycles no P2021-10-007), and the Top Level Research Area SusInnoScience of the federal state of Rheinland-Pfalz.

**Author contributions**

H.-J.B. and M.K. supervised the work. X.B., X.G. and Y.X. proposed the work. X.B., Y.X., X.L., M.K. and H.-J.B. made an experimental plan. X.B. and X.Z. prepared the samples and performed the experiments. D.G. performed the electrochemical measurements. X.B., M.K., C.H., Y.X. and R.B. designed the setup. X.B., A.D.R., X.Z., M.K., H.-J.B., Q.L., Y.X., and X.L. analyzed the data and prepared the manuscript. All authors reviewed the manuscript.